\documentclass[prd,twocolumn,twoside,preprintnumbers,superscriptaddress,nofootinbib]{revtex4}

\pdfoutput=1
\usepackage{amsmath,amssymb,bm,graphicx,bbold,epsf,colordvi}
\usepackage{lipsum}
\usepackage{braket}
\allowdisplaybreaks 
\usepackage{color}
\newcommand\f{\frac}
\newcommand\as{\alpha_s}
\newcommand{\ba}{\begin{eqnarray}}
\newcommand{\ea}{\end{eqnarray}}
\newcommand{\be}{\begin{equation}}
\newcommand{\ee}{\end{equation}}
\newcommand{\nn}{\nonumber}

\def\OMIT#1{{}}

\def\SCETG{${\rm SCET}_{\rm G}\,$}

\begin{document}

\preprint{\vbox{\hbox{LA-UR-17-20230}}}

\title{Inclusive production of small radius jets in heavy-ion collisions}

\author{Zhong-Bo Kang}
\affiliation{Department of Physics and Astronomy, University of California, Los Angeles, CA 90095, USA}
\affiliation{Mani L. Bhaumik Institute for Theoretical Physics, University of California, Los Angeles, CA 90095, USA}
\affiliation{Theoretical Division, Los Alamos National Laboratory, Los Alamos, NM 87545, USA}
\author{Felix Ringer}
\affiliation{Theoretical Division, Los Alamos National Laboratory, Los Alamos, NM 87545, USA}
\author{Ivan Vitev}
\affiliation{Theoretical Division, Los Alamos National Laboratory, Los Alamos, NM 87545, USA}

\begin{abstract}
We develop a new formalism to describe the inclusive production of small radius jets in heavy-ion collisions, which is consistent with jet calculations in the simpler proton-proton system. Only at next-to-leading order (NLO) and beyond, the jet radius parameter $R$ and the jet algorithm dependence of the jet cross section can be studied and a meaningful comparison to experimental measurements is possible. We are able to consistently achieve NLO accuracy by making use of the recently developed semi-inclusive jet functions within Soft Collinear Effective Theory (SCET). In addition, single logarithms of the jet size parameter $\alpha_s^n\ln^n R$ are resummed to next-to-leading logarithmic (NLL$_R$) accuracy. The medium modified semi-inclusive jet functions are obtained within the framework of SCET with Glauber gluons that describe the interaction of jets with the medium. We present numerical results for the suppression of inclusive jet cross sections in heavy ion collisions at the LHC and the formalism developed here can be extended directly to corresponding jet substructure observables. 
\end{abstract}

\maketitle
\section{Introduction \label{sec:introduction}}

In heavy-ion collisions at RHIC and LHC, a quark-gluon plasma (QGP) can be created and studied by using both hard and soft probes~\cite{Wang:2016opj}. Jets are produced in hard-scattering events and constitute one of most frequently studied examples of hard probes in heavy-ion collisions. Jets traverse the hot and dense QCD medium and are identified as energetic and collimated sprays of particles in the detectors. Examination of their properties can, therefore, provide information about the QGP. The LHC experimental collaborations CMS~\cite{Khachatryan:2016jfl}, ATLAS~\cite{Aad:2012vca} and ALICE~\cite{Abelev:2013fn} have provided precise data sets for the inclusive production of jets in both proton-proton and heavy-ion collisions. In heavy-ion collisions jets are modified, or quenched, due to the interaction with the QCD medium. Most commonly, the quenching of jet production yields is studied using the nuclear modification factor $R_{AA}$, which is given by the ratio of the respective cross section in heavy-ion collisions normalized by the corresponding proton-proton baseline. In order to reliably extract information about the QGP from the available data sets, it is important that the experimentally achieved precision is matched with corresponding theoretical calculations. This is precisely what we are going to address in this work.

The identification of jets relies on a jet algorithm that specifies when particles are clustered together into the same jet. Typical algorithms used by the experimental analyses involve for example the anti-$k_T$ and the cone algorithm~\cite{Sterman:1977wj,Cacciari:2008gp}. In addition, jets are defined by the jet parameter $R$ which represents the size of the identified jets, see e.g.~\cite{Soyez:2008su,Mukherjee:2012uz} for more details. The first non-trivial order in the perturbative expansion of the cross section where these specifications play a role is next-to-leading order (NLO) in QCD. Therefore, full NLO is the minimally required perturbative order allowing meaningful comparisons between theory and the experimental measurements. In addition, parton distribution functions (PDFs) are fitted at NLO, typically for larger values of $R\sim 0.7$. For the analyses of heavy-ion collisions, however, the jet size parameter is typically chosen to be relatively small, $R\sim 0.2-0.4$, in order to minimize fluctuations in the heavy-ion background. The perturbative series exhibits a single logarithmic structure $\alpha_s^n\ln^n R$ to all orders in the QCD strong coupling constant, which have to be resummed to render the convergence of the perturbative calculations. 

Such a $\ln R$-resummation has recently been achieved for proton-proton collisions to next-to-leading logarithmic (NLL$_R$) accuracy~\cite{Kang:2016mcy} within the framework of Soft Collinear Effective Theory (SCET)~\cite{Bauer:2000yr,Bauer:2001ct,Bauer:2001yt,Beneke:2002ph}. See~\cite{Dasgupta:2014yra,Dasgupta:2016bnd,Kaufmann:2015hma,Dai:2016hzf} for related work along these lines. 
It was also demonstrated that the cross section for inclusive jets can be factorized into convolution products of PDFs, hard functions and so-called semi-inclusive jet functions  $J_i(z,\omega_J R,\mu)$ (siJFs)~\cite{Kang:2016mcy}. The siJFs describe the formation of a jet with energy $\omega_J$ and jet parameter $R$ originating from a parent parton $i$ at scale $\mu$. They satisfy the same timelike DGLAP equations that govern the scale evolution of fragmentation functions. By solving the DGLAP equations, the resummation of logarithms $\ln R$ can be achieved. For values of $R$ in the range of $0.2-0.4$, fixed NLO calculations fail to describe the experimental data~\cite{Khachatryan:2016jfl}. When the resummation of $\ln R$ terms is included, good agreement can be achieved, as we will show below. The need of $\ln R$ resummation to describe the proton-proton baseline makes this the ideal starting point to also study inclusive jet production in heavy-ion collisions. In this work, we extend our earlier calculations for proton-proton collisions to heavy-ion collisions. See for example~\cite{Vitev:2008rz,He:2011pd,Muller:2012zq,Zapp:2013zya,Armesto:2015ioy,Chien:2015hda,Chang:2016gjp,Wang:2016fds} for earlier work on the description of jets in heavy-ion collisions.

\begin{figure*}[t]
\includegraphics[width=0.85\textwidth]{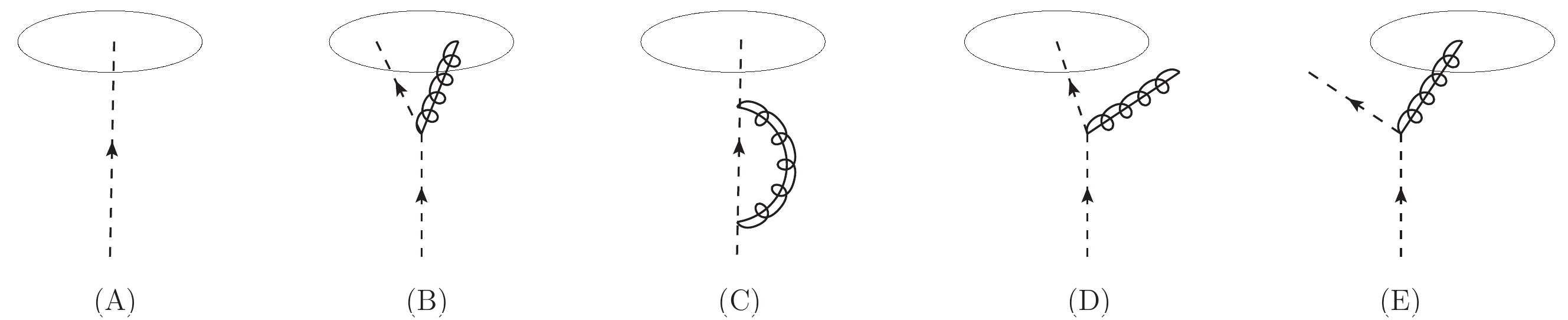}
\caption{Feynman diagrams for quark initiated jets. (A) is the leading-order contribution and (B)-(E) are the relevant diagrams at NLO: (B) both partons are inside the jet, (C) virtual correction, (D) and (E) only one parton is inside the jet. The dashed lines correspond to a collinear quarks and the curly lines to collinear gluons within SCET. \label{fig:vac}}
\end{figure*}

We address the medium modification within the effective field theory (EFT) framework of SCET with Glauber gluons which is generally denoted by \SCETG~\cite{Idilbi:2008vm,Ovanesyan:2011xy}. The interaction of collinear quarks and gluons with the hot and dense QCD medium can be described via the exchange of Glauber gluons. Within \SCETG, the relevant interaction terms are included at the level of the Lagrangian. By making use of the collinear sector of the corresponding EFT, the full collinear in-medium splitting functions have been derived in the past years to first order in opacity~\cite{Ovanesyan:2011kn,Fickinger:2013xwa,Ovanesyan:2015dop,Kang:2016ofv}. When finite quark masses are neglected, the in-medium splitting functions are given by the vacuum splitting functions times a modification factor that depends on the properties of the medium. The opacity expansion for the medium interactions is analogous to the traditional Gyulassy-Levai-Vitev (GLV) approach to parton energy loss in the QCD medium~\cite{Gyulassy:2000fs}. At first order in opacity, an average number of uncorrelated interactions with the medium is taken into account. Higher orders in the opacity expansion correspond to correlations between the interactions which are yet to be calculated and are neglected in this work. In the traditional GLV approach, all radiated gluons in the splitting processes are approximated to be soft. Within \SCETG, one can go beyond this approximation and obtain full control of the collinear dynamics of splitting processes in the medium. For example, the in-medium splitting functions have already been successfully applied to describe the modification of light hadrons~\cite{Kang:2014xsa,Chien:2015vja} as well as heavy flavor mesons~\cite{Kang:2016ofv} in heavy-ion collisions. 

In this work, we derive an analogous treatment of the in-medium effects for inclusive jets by defining in-medium siJFs. In the vacuum, the siJFs can be written in terms of collinear vacuum splitting functions. In the medium, we need to include additional contributions to the siJFs that can be expressed in terms of collinear in-medium splitting functions derived from \SCETG. In general, both radiative and collisional energy loss can play a role in modifying jet production in the medium. In this work, we concentrate on the high-$p_T$ jets and thus only consider radiative energy loss, and will leave collisional energy loss for future publications. In addition, we include Cold Nuclear Matter (CNM) effects. 

The remainder of this paper is organized as follows. In Section~\ref{sec:theory}, we recall the basic framework for inclusive jet production in proton-proton collisions for small-$R$ jets. We outline a consistent extension of the jet cross section to heavy-ion collisions using the in-medium collinear splitting functions obtained within \SCETG. In Section~\ref{sec:numerics}, we present numerical results for the nuclear modification factor $R_{AA}$ and we compare to recent data from the LHC. Finally, we draw our conclusions in Section~\ref{sec:conclusions}.

\section{Theoretical Framework \label{sec:theory}}

We start by summarizing the main results of~\cite{Kang:2016mcy} for inclusive jet production in proton-proton collisions. We then outline how this framework can be 
extended to the heavy-ion case. 

\subsection{Proton-proton collisions}
The factorized form of the double differential cross section for inclusive jets with a given transverse momentum $p_T$ and rapidity $\eta$ is given by
\be\label{eq:factorization}
\f{d\sigma^{pp\to \mathrm{jet}X}}{dp_Td\eta}=\sum_{a,b,c} f_a\otimes f_b\otimes H_{ab}^c\otimes J_c \, .
\ee
Here, we suppressed the arguments of the various functions for better readability. See~\cite{Kang:2016mcy} for more details. The symbols $\otimes$ denote convolution products and we are summing over all relevant partonic channels. The $f_{a,b}$ denote the PDFs, $H_{ab}^c$ are hard-functions and the $J_c$ are the siJFs. The hard-functions are evaluated up to NLO and were shown~\cite{Kang:2016mcy} to be the same as the hard-functions for inclusive hadron production $pp\to hX$, see~\cite{Aversa:1988vb,Jager:2002xm}. Note that Eq.~(\ref{eq:factorization}) is a factorization of purely hard-collinear dynamics, i.e. no soft function is needed. The siJFs are perturbatively calculable functions that describe the formation of the observed jet originating from a parent parton.

If both $H_{ab}^c$ and $J_c$ are expanded to NLO, we get back to the standard NLO results for inclusive jets as derived in~\cite{Aversa:1989xw,Jager:2004jh,Mukherjee:2012uz}. Within the framework developed in~\cite{Kang:2016mcy}, we can go beyond the fixed order approach. Using the siJFs in Eq.~(\ref{eq:factorization}) represents an additional final state factorization. As pointed out in~\cite{Dasgupta:2014yra,Dasgupta:2016bnd}, fixed order jet cross sections can have a vanishing, unphysical scale dependence. This problem is overcome by using the factorized form of the cross section in Eq.~(\ref{eq:factorization}), where the interpretation of the QCD scale uncertainty as a measure of missing higher order corrections is restored. See~\cite{Kang:2016mcy} for numerical results concerning the scale dependence.

In~\cite{Kang:2016mcy}, the siJFs $J_i(z,\omega_J R,\mu)$ were calculated to NLO from their operator definition within SCET. We have $z=\omega_J/\omega$, where $\omega_J$ ($\omega$) denotes the jet (initiating parton) energy. Here, we summarize only the results for quark initiated jets to keep the discussion short. Up to NLO, one has to consider the diagrams presented in Fig.~\ref{fig:vac} for jets that are initiated by an outgoing quark. (A) corresponds to the leading-order (LO) diagram. To LO, one finds $J_q^{(0)}(z,\omega_J R,\mu)=\delta(1-z)$. At NLO, one has to consider the contributions (B)-(E). Here (B) corresponds to a splitting process, where both final state partons are in the jet. (C) is a virtual correction and (D), (E) are the contributions, where one of the final state partons exits the jet. All contributions (B)-(E) can be written in terms of integrals over the quark-to-quark or quark-to-gluon LO Altarelli-Parisi splitting functions~\cite{Ritzmann:2014mka}. We make use of this fact below when deriving the in-medium siJFs. For completeness, we present here the result for the quark siJF in dimensional regularization up to NLO for the anti-$k_T$ algorithm
 \begin{align}
 \label{eq:siJFNLO}
 J_q(z,\omega R,\mu)  = & \delta(1-z)\nn\\ 
 &\hspace*{-2.2cm}+\f{\as}{2\pi}\left(\f{1}{\epsilon}+\ln\left(\f{\mu^2}{\omega^2\tan^2(R/2)}\right)-2\ln z\right)\left[P_{qq}(z)+P_{gq}(z)\right]\nn \\
 &\hspace*{-2.2cm}-\f{\as}{2\pi}\Bigg\{C_F\left[2(1+z^2)\left(\f{\ln(1-z)}{1-z}\right)_++(1-z)\right]\nn\\
&\hspace*{-2.2cm} -\delta(1-z) C_F\left(\f{13}{2}-\f{2\pi^2}{3}\right)+2 P_{gq}(z) \ln(1-z)+C_F z\Bigg\} \,,
 \end{align}
where $P_{qq}(z)$ and $P_{gq}(z)$ are the usual LO Altarelli-Parisi splitting functions. The remaining $1/\epsilon$ pole is a UV pole which is removed by renormalization. The associated RG equations turn out to be the same DGLAP evolution equations that are also satisfied by fragmentation functions, which describe the transition of a final state parton into a specific observed hadron. In other words, the DGLAP equations for the siJFs read
\be\label{eq:DGLAP}
\mu\f{d}{d\mu}J_i=\f{\as}{2\pi}\sum_j P_{ji}\otimes J_j \,,
\ee
where we omitted again the arguments of the involved functions. These evolution equations can be solved in Mellin moment space using the methods developed in~\cite{Vogt:2004ns,Anderle:2015lqa}. By solving the DGLAP equations, we obtain the evolved siJFs as they appear in the factorization theorem in Eq.~(\ref{eq:factorization}). The resummation of terms $\sim\ln R$ is achieved by choosing $\mu\sim \omega_J\tan(R/2)\approx p_T R$ in Eq.~(\ref{eq:siJFNLO}) which eliminates terms $\sim\ln R$ in the fixed order result. We then evolve the siJFs through the DGLAP equations from this characteristic scale to the hard scale $\mu\sim p_T$. In section~\ref{sec:numerics}, we present some numerical results at NLO+NLL$_R$ accuracy showing the impact of the $\ln R$ resummation for narrow jets.

Note that in~(\ref{eq:siJFNLO}), we chose to express the result in terms of the initiating parton energy $\omega=\omega_J/z$. This convention differs from the one chosen in~\cite{Kang:2016mcy}. The different convention results in an additional term $\sim -2\ln z$ in the second line in of Eq.~(\ref{eq:siJFNLO}). In principle, $\omega_J$ is the relevant external quantity as it is related to the observed jet transverse momentum $\omega_J=2p_T\cosh \eta $. However, the underlying structure of the siJFs becomes more apparent when writing the result in terms of the energy $\omega$ of the initiating parton. This will be particularly relevant for deriving the in-medium siJFs below. Also note that the sum rule for the siJFs associated with momentum conservation of the initiating parton $i$,
\be
\int_0^1 dz\, z J_i(z,\omega R,\mu)=1\,,
\ee
(anti-$k_T$ algorithm) only holds when expressing the siJFs in terms of the parton energy $\omega$ instead of the jet energy $\omega_J$, see also~\cite{Dai:2016hzf}.

\subsection{Heavy-ion collisions}
We now turn to the cross section for inclusive jets produced in heavy-ion collisions. First, we note that the QGP is only present in the final state after the hard-scattering event. Therefore, it is sufficient to modify only the siJFs which capture the formation of the observed jet. Second, all Feynman diagrams shown in Fig.~\ref{fig:vac} that are relevant for the vacuum also appear in the medium calculation. In other words, in the heavy ion collisions, the siJFs obtained in proton-proton collisions are modified as
\be\label{eq:sum}
J_i \to J_i^{\mathrm{vac}}+J_i^{\mathrm{med}} \, ,
\ee
where $J_i^{\mathrm{vac}}$ are the vacuum contributions, and $J_i^{\mathrm{med}}$ are the in-medium siJFs that take into account medium interactions. 

\begin{figure}[htb]
\vspace*{.7cm}
\includegraphics[width=220pt]{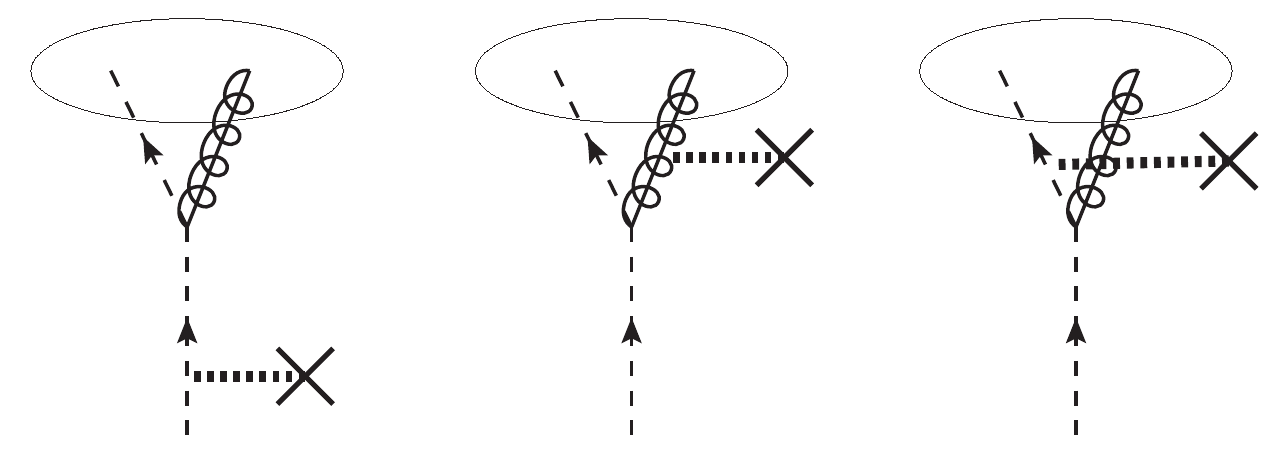}
\caption{Single-Born diagrams contributing to the medium siJF $J_q^{\mathrm{med}}$ where both partons are inside the jet, cf. Fig.~\ref{fig:vac} (B). The dotted lines represent the interaction with the QCD medium via Glauber gluon exchange. \label{fig:med}}
\end{figure}

The Feynman diagrams that contribute to $J_i^{\mathrm{med}}$ can be obtained from the corresponding vacuum ones shown in Fig.~\ref{fig:vac}. As an example, see Fig.~\ref{fig:med} for the relevant so-called single-Born (SB) diagrams ${\cal A}^{\mathrm{med}}_{\mathrm{SB}}$ for the case that both partons remain inside the jet (B). In addition, we need to calculate double-Born (DB) diagrams ${\cal A}^{\mathrm{med}}_{\mathrm{DB}}$ where two interactions with the medium are considered. The relevant diagrams are not shown here explicitly. In order to obtain a physical in-medium cross section, we schematically need to calculate the combination
\be
|{\cal A}^{\mathrm{med}}_{\mathrm{SB}}|^2+2\mathfrak{Re}\left\{{\cal A}^{\mathrm{med}}_{\mathrm{DB}}\times {\cal A}^{\mathrm{vac}}\right\}\,,
\ee
where ${\cal A}^{\mathrm{vac}}$ denotes the vacuum diagrams as shown in Fig.~\ref{fig:vac}. See~\cite{Ovanesyan:2011xy,Kang:2016ofv} for a more detailed discussion. 

As mentioned above, in the calculation of the vacuum siJFs $J_i^{\mathrm{vac}}$, all contributions (B)-(E) in Fig.~\ref{fig:vac} can be expressed in terms of integrals over the LO real emission splitting functions $P_{ji}(z,q_\perp)$. For example, for a quark-to-quark splitting, we have
\be\label{eq:Pqq}
P_{qq}(x,q_\perp) = \f{\as}{\pi} C_F\f{1+x^2}{1-x}\f{1}{q_\perp} \,,
\ee
where we include explicitly a $1/q_\perp$ (its transverse momentum dependence) relative to the Altarelli Parisi splitting functions used above. See Ref.~\cite{Kang:2016mcy} for more details. In order to obtain the corresponding in-medium siJFs $J_i^{\mathrm{med}}$, the vacuum splitting functions $P_{ji}(z,q_\perp)$ are replaced with the collinear in-medium splitting functions $P_{ji}^{\mathrm{med}}(z,q_\perp)$ derived from \SCETG. With all quark masses set to zero, one finds that they have the following structure
\be\label{eq:Pjimed}
P_{ji}^{\mathrm{med}}(z,q_\perp) = P_{ji}(z,q_\perp)\; f_{ji}(z,q_\perp; \beta)
\ee
where the characteristics (properties) of the medium are collectively denoted by $\beta$~\cite{Kang:2014xsa}. The functions $f_{ji}(z,q_\perp; \beta)$ describe the modification of the vacuum splitting function due to the presence of the QCD medium. 

The \SCETG splitting functions can be partly evaluated analytically. The explicit form of the relevant \SCETG in-medium splitting functions for massless partons can be found for example in~\cite{Ovanesyan:2011kn}. Eventually, they have to be integrated also over the size of the medium and the transverse momentum transfer that is acquired due to the medium interactions. These integrations can only be evaluated numerically, as they depend on the specific properties of the medium. Therefore, we have to find a way to evaluate the in-medium siJFs $J_i^{\mathrm{med}}$ numerically, such that all divergences that appear only at the intermediate steps of the calculation cancel. For the calculation of the in-medium siJFs, we choose to work in a cut-off scheme rather than dimensional regularization in order to facilitate the numerical evaluation. A single regulator $\mu$ cutting off UV divergences is sufficient as there is only one remaining UV divergence once the contributions from all diagrams are taken into account. This can also be seen from the vacuum result of the quark siJF in dimensional regularization in Eq.~(\ref{eq:siJFNLO}), where there is only one $1/\epsilon$ UV pole left. Note that in dimensional regularization, the virtual correction as shown in Fig.~\ref{fig:vac} (C) leads to a scaleless integral. Effectively, it only changes IR poles into UV poles. However, in order to work out the in-medium result numerically, we have to explicitly take this contribution into account.

\begin{figure*}[t]
\includegraphics[width= 0.75 \textwidth]{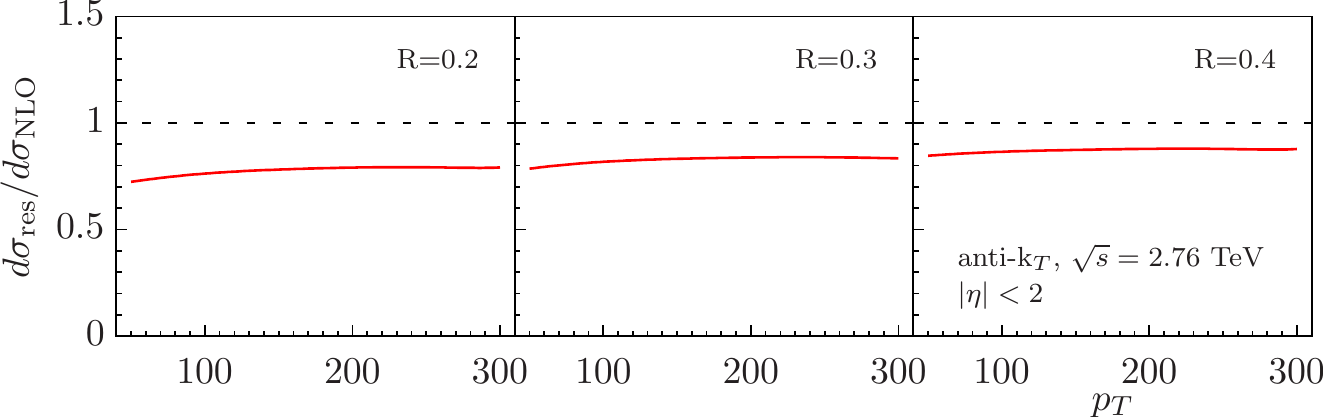}
\caption{Ratio of the resummed NLO+NLL$_R$ inclusive jet cross section and the fixed NLO result in proton-proton collisions. The kinematics are chosen as in the CMS analysis of~\cite{Khachatryan:2016jfl}: $\sqrt{s}=2.76$~TeV, $|\eta|<2$ and the observed jets are reconstructed using the anti-$k_T$ algorithm. The jet size parameter is chosen as $R=0.2,\,0.3,\,0.4$ in the three panels from left to right.  \label{fig:vac-pp}}
\end{figure*}

We now present the expressions for the individual contributions (B)-(E) to the quark siJF written in terms of integrals over splitting functions. Here, we generally write the splitting functions as $P_{ji}(z,q_\perp)$. For $J_i^{\mathrm{vac}}$, we refer to the vacuum ones, see e.g. Eq.~(\ref{eq:Pqq}), and for $J_i^{\mathrm{med}}$ to the in-medium \SCETG splitting functions as in Eq.~(\ref{eq:Pjimed}). The two cases eventually will be summed over as indicated in Eq.~(\ref{eq:sum}). Following~\cite{Kang:2016mcy}, we can write the result associated with diagram (B), where both partons are inside the jet, as the following integral over the quark-to-quark splitting function~\cite{Ellis:2010rwa,Kang:2016mcy}
\be\label{eq:B}
(\mathrm{B})=\delta(1-z)\int_0^1 dx \int_0^{x(1-x)\omega\tan(R/2)} dq_\perp P_{qq}(x,q_\perp) \, .
\ee
Note that here we can use the energy of the initiating parton or the jet as the delta function enforces $z=1$ or equivalently $\omega=\omega_J$. We would like to point out that the result for $(\mathrm{B})$ by itself is divergent both in the vacuum and in the medium. This can be most easily seen for the vacuum case using dimensional regularization, where one finds both double poles $1/\epsilon^2$ and single poles $1/\epsilon$~\cite{Kang:2016mcy}. 

In order to evaluate the in-medium result numerically, we first have to appropriately combine the result in Eq.~(\ref{eq:B}) with the other contributions (C)-(E). We continue with the virtual correction shown in Fig.~\ref{fig:vac} (C). Following~\cite{Collins:1988wj}, this contribution can be written as
\be
(\mathrm{C})=-\delta(1-z)\int_0^1 dx \int_0^\mu dq_\perp P_{qq}(x,q_\perp) \,,
\ee
where we introduced the UV cut-off $\mu$ as the upper integration boundary of the $q_\perp$ integral. Both contributions (B) and (C) are $\sim\delta(1-z)$ and we can directly combine them as
\be\label{eq:BC}
\mathrm{(B)+(C)}=\delta(1-z)\int_0^1 dx \int^\mu_{x(1-x)\omega\tan(R/2)} dq_\perp P_{qq}(x,q_\perp) \, .
\ee
The contribution from the diagram (D), where the gluon exits the jet can be written as
\be\label{eq:D}
(\mathrm{D})= \int_{z(1-z)\omega\tan(R/2)}^\mu dq_\perp P_{qq}(z,q_\perp) \,,
\ee
where symmetry of the lower integration boundary with respect to $z\leftrightarrow 1-z$ is obtained by using the energy $\omega$ instead of $\omega_J$. 
Note that the contribution (D) in Eq.~(\ref{eq:D}) by itself is divergent for $z\to 1$ as $P_{qq}(z,q_\perp)\sim 1/(1-z)$. Similarly, the result for (B) + (C) in Eq.~(\ref{eq:BC}) is divergent by itself. However, it can be seen immediately that the two results have a very similar structure and they can be combined together by introducing a plus distribution:
\begin{align}\label{eq:one-loop}
\mathrm{(B)+(C)+(D)} =  \left[ \int_{z(1-z)\omega\tan(R/2)}^\mu dq_\perp P_{qq}(z,q_\perp) \right]_+,
\nonumber
\end{align}
where the plus distribution is defined as usual via
\be
\int_0^1 dz\, f(z)[g(z)]_+\equiv \int_0^1dz \,(f(z)-f(1))g(z)\, .
\ee
Apparently such a combination written in this form is finite for $z\to 1$. 

Finally, for the case that the gluon makes the jet and the quark is outside of the jet, we have
\be
(\mathrm{E})= \int_{z(1-z)\omega\tan(R/2)}^\mu dq_\perp P_{gq}(z,q_\perp) \,.
\ee
The splitting function $P_{gq}(x,q_\perp)$ describes the quark-to-gluon splitting process and can be obtained from Eq.~(\ref{eq:Pqq}) by substituting $x\to 1-x$. Diagram (E) is finite by itself for $z\to 1$. When summing over all diagrams at NLO (B)-(E), we can write the result as
\begin{align}\label{eq:one-loop}
J^{\mathrm{med},(1)}_q(z,\omega R,\mu)  = & \left[ \int_{z(1-z)\omega\tan(R/2)}^\mu dq_\perp P_{qq}(z,q_\perp) \right]_+ \nn \\
& \hspace{-7mm} +  \int_{z(1-z)\omega\tan(R/2)}^\mu dq_\perp P_{gq}(z,q_\perp) \,.
\end{align}
The result for $J_q^{\mathrm{med}}$ written in this form is finite for $z\to 1$ and we only need the upper UV cut-off $\mu$, which is suitable for numerical implementations and integrations.

The structure of the result in Eq.~(\ref{eq:one-loop}) is completely analogous to the sum of real and virtual NLO corrections to the partonic calculation of fragmentation functions (FFs), see~\cite{Collins:1988wj,Kang:2016ofv}. Note that for jet production the sum of diagrams (B) + (C) in Eq.~(\ref{eq:BC}) is analogous to the virtual NLO correction for FFs. On the other hand, the results from diagrams (D) and (E) involving radiation outside of the jet corresponds to the real correction for FFs. This close analogy allows us to treat the medium modification of hadron and jet production yields in heavy-ion collisions essentially on the same footing. The result for the gluon siJF can be obtained in an analogous way. However, there are some subtleties when introducing the plus prescription. The same issues were discussed in details in~\cite{Kang:2016ofv} in the context of heavy flavor production in heavy-ion collisions.

Finally, following the additive structure of the vacuum and in-medium siJFs described in Eq.~(\ref{eq:sum}), we obtain the following structure of the inclusive jet production cross section in heavy-ion (PbPb) collisions
\be\label{eq:mediumresult}
d\sigma^{\mathrm{jet}}_{\mathrm{PbPb}} = d\sigma_{pp}^{\mathrm{jet,vac}}+d\sigma_{\mathrm{PbPb}}^{\mathrm{jet,med}} \,.
\ee
Here the first term is given in Eq.~(\ref{eq:factorization}). The second term is given by
\be
d\sigma_{\mathrm{PbPb}}^{\mathrm{jet,med}}= \sum_{i=q,\bar q,g} \sigma_i^{(0)}\otimes J_i^{\mathrm{med}} \,,
\ee
where $\sigma_i^{(0)}$ is the LO production cross section for quarks and gluons since $J_i^{\mathrm{med}}$ is formally already NLO. Ideally, one would like to evolve the in-medium siJFs also using the DGLAP evolution equations in~(\ref{eq:DGLAP}). However, in this work we limit ourselves to performing the $\ln R$ resummation only for the first term in~(\ref{eq:mediumresult}) and we add the second term consistently at NLO.

\section{Numerical Results \label{sec:numerics}}
In this section, we present numerical results for inclusive jet production at NLO+NLL$_R$ accuracy in  both proton-proton and heavy-ion collisions using the framework outlined in section~\ref{sec:theory}. For all numerical results presented in this section, we use the CT14 NLO set of PDFs~\cite{Dulat:2015mca}. 

We start by considering inclusive small-$R$ jet production in proton-proton collisions at the LHC. In~\cite{Kang:2016mcy}, we presented a more detailed study of the effects of $\ln R$ resummation and we also considered the scale dependence of the cross section. Only very recently, the CMS collaboration presented precise measurements of small-$R$ jets~\cite{Khachatryan:2016jfl} where the jet radius parameter was chosen as $R=0.2,\,0.3,\,0.4$. The jets are reconstructed using the anti-$k_T$ algorithm at $\sqrt{s}=2.76$~TeV and $|\eta|<2$. It was shown in~\cite{Khachatryan:2016jfl} that standard NLO calculations are not able to describe the data. The discrepancy becomes larger for small $R$ and low $p_T$. One has to keep in mind that modern sets of PDFs are also fitted to inclusive jet spectra which constrain, in particular, the gluon PDF at large $x$. However, the jet size parameter $R$ is typically relatively large ($\sim 0.7$) for the data sets included in these fits. In Fig.~\ref{fig:vac-pp}, we show results when the resummation of $\ln R$ terms is included. We choose the same kinematics as in the CMS analysis and plot the ratio of the NLO+NLL$_R$ results normalized by the NLO result for the three different values of $R$. By comparing with the results shown in~\cite{Khachatryan:2016jfl}, we find that the discrepancy between the theory calculation at NLO and the data for small-$R$ jets is well described when $\ln R$ resummation is included. We conclude that $\ln R$ resummation is necessary to describe small-$R$ jet data at the LHC. It is, therefore, desirable to work with this proton-proton baseline calculation when considering the modification of inclusive jet spectra in heavy-ion collisions.
 \begin{figure}[hbt]
\includegraphics[width= 210pt]{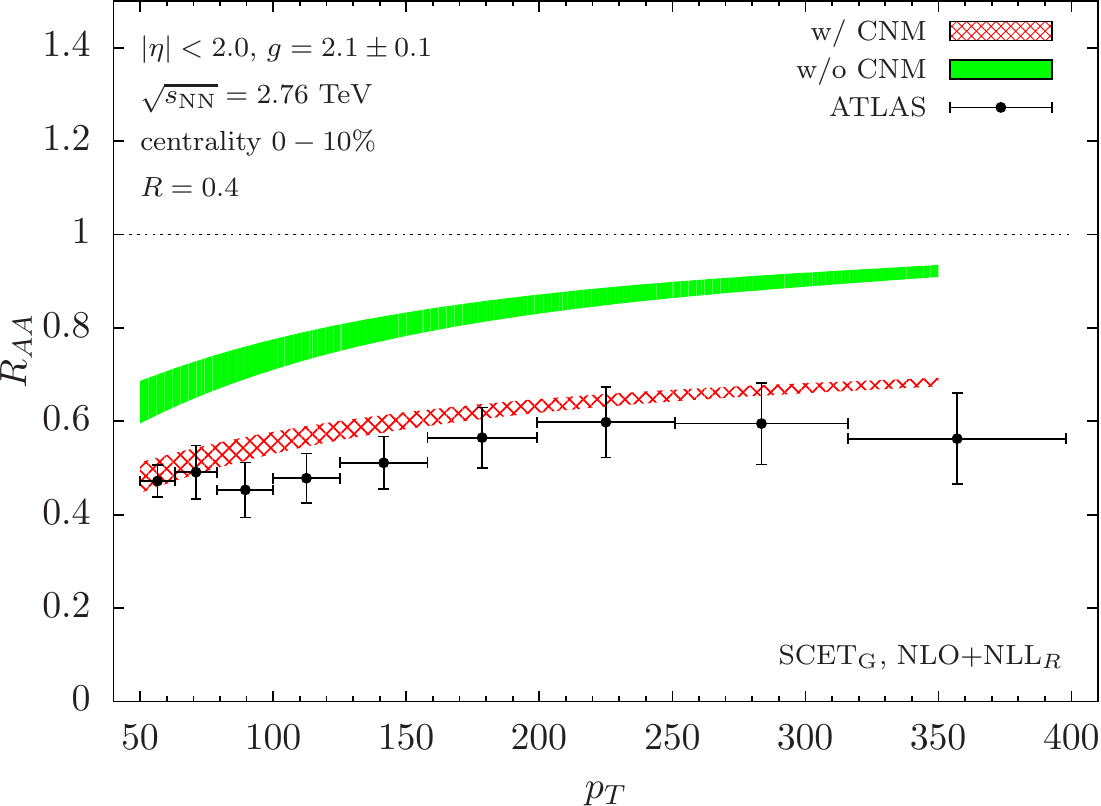}
\caption{The nuclear modification factor $R_{AA}$ for heavy-ion collisions at the LHC for $\sqrt{s_{\mathrm{NN}}}=2.76$~TeV, $|\eta|<2$, $R=0.4$, anti-$k_T$ jets and $0-10\%$ centrality. The \SCETG results at NLO+NLL$_R$ are shown without CNM effects (green band) and with CNM effects (hatched red band). The coupling strength between the jet and the QCD medium is chosen as $g=2.1\pm 0.1$. We compare to the ATLAS data of~\cite{Aad:2012vca}. Statistical and systematic errors are added in quadrature.  \label{fig:ATLAS}}
\end{figure}

Having established a framework that can describe the proton-proton baseline, we now turn to numerical results for inclusive jet production in heavy-ion collisions. We consider the nuclear modification factor $R_{AA}$ which is defined as
\be
R_{AA}=\f{d\sigma^{\mathrm{PbPb}\to\mathrm{jet}X}}{\braket{N_{\mathrm{coll}}} d\sigma^{pp\to\mathrm{jet}X}} \,,
\ee
where $\braket{N_{\mathrm{coll}}}$ is the average number of binary nucleon-nucleon collisions. We start by assessing the numerical impact of possible CNM effects and by comparing our results to the ATLAS data of~\cite{Aad:2012vca} in Fig.~\ref{fig:ATLAS}. The ATLAS data for the nuclear modification factor $R_{AA}$ was taken at $\sqrt{s_{\mathrm{NN}}}=2.76$~TeV and $0-10\%$ centrality for $|\eta|<2$. The jets were reconstructed using the anti-$k_T$ clustering algorithm with a jet size parameter of $R=0.4$. In this figure, the statistical and systematic errors of the ATLAS data were added in quadrature. We leave the coupling strength $g$ between the jet and the medium as a parameter that can eventually be determined by comparing to data. We choose $g=2.1$ as a central value and we obtain a band as shown in Fig.~\ref{fig:ATLAS} by varying that value of $g$ by $\pm 0.1$. Note that the coupling $g$, and the associated $\alpha_s=g^2/(4\pi)$, correspond to the vertices shown in Fig.~\ref{fig:med} between the Glauber gluons (dotted lines) and the collinear partons. The green band in Fig.~\ref{fig:ATLAS} shows our \SCETG results at NLO+NLL$_R$ accuracy without CNM effects. We emphasize again that resummed accuracy is achieved for the vacuum contribution. The medium contribution is consistently included at NLO. The hatched red band shows the same results but with CNM effects. We obtain a good description of the ATLAS data once CNM effects are included. The initial state energy loss considered here corresponds to a momentum exchange scale of $\mu_{\mathrm{CNM}}=0.35$~GeV. See~\cite{Vitev:2007ve} for more details. As mentioned above, we do not include collisional energy loss in this work. Eventually, it is important to clearly disentangle the numerical size of the different contributions of radiative and collisional energy loss mechanisms as well as CNM effects. One possibility is to consider jet substructure observables where CNM effects are expected to only play a marginal role~\cite{Chien:2015hda}. More detailed studies along these lines will be left for future work.
 \begin{figure*}[t]
\includegraphics[width= 210pt]{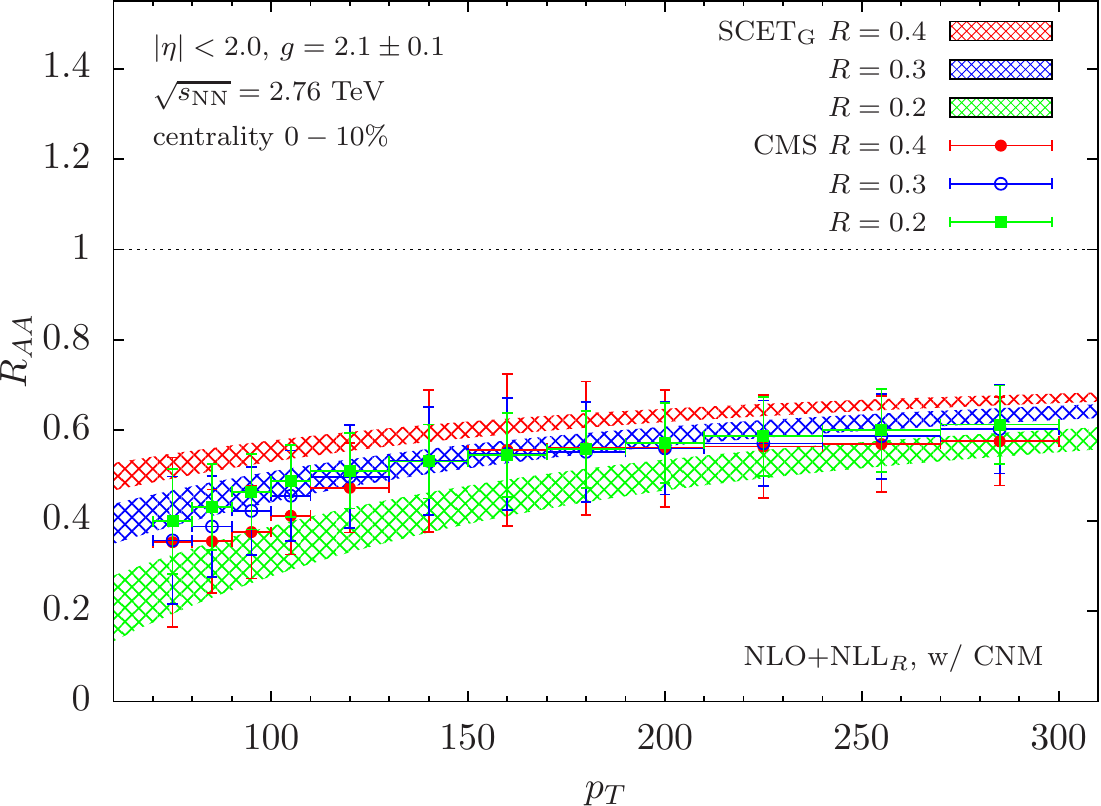} \hspace*{0.2in}\includegraphics[width= 210pt]{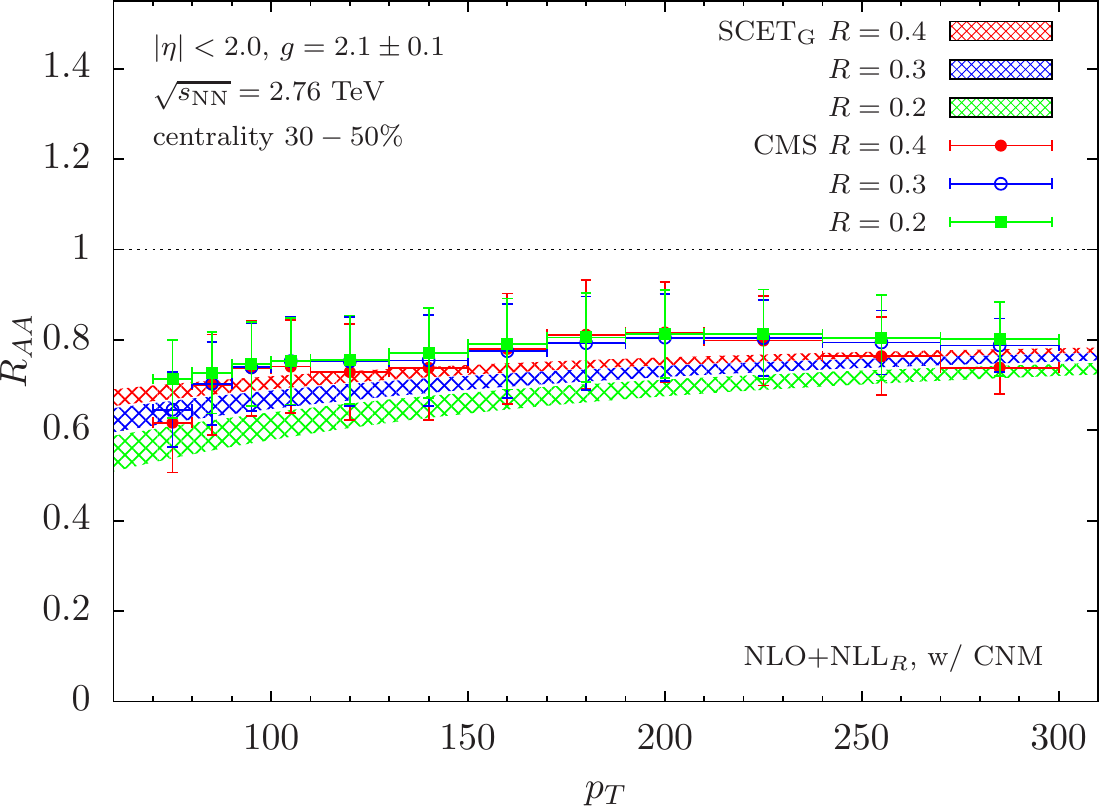}
\caption{Comparison of the \SCETG results at NLO+NLL$_R$ with the CMS data of~\cite{Khachatryan:2016jfl} for $R=0.2$ (green), $R=0.3$ (blue) and $R=0.4$ (red). Statistical and systematic errors are added in quadrature. We have $\sqrt{s_{\mathrm{NN}}}=2.76$~TeV, $|\eta|<2$ and we choose the coupling constant of the medium to the jet as $g=2.1\pm 0.1$. On the left side, we show the results for central collisions ($0-10\%$ centrality) and on the right side for mid-peripheral collisions ($30-50\%$ centrality).  \label{fig:CMS}}
\end{figure*}

Finally, we compare our calculations to the CMS measurement of the nuclear modification factor $R_{AA}$~\cite{Khachatryan:2016jfl} in Fig.~\ref{fig:CMS}. Similar to the ATLAS measurement, the data was taken at $\sqrt{s_{\mathrm{NN}}}=2.76$~TeV for $|\eta|<2$ using the anti-$k_T$ algorithm. We show the results for both central collisions ($0-10\%$ centrality) on the left side as well as for mid-peripheral collisions ($30-50\%$ centrality) on the right side of Fig.~\ref{fig:CMS}. The results for different values of the jet radius parameter are all shown in one plot: $R=0.2$ (green), $R=0.3$ (blue) and $R=0.4$ (red). Again, systematic and statistical errors are added in quadrature. We show our \SCETG based results at NLO+NLL$_R$ accuracy for the different values of $R$ using the same color coding. We only show our results with CNM effects and we choose the coupling of the jet and the medium as $g=2.1\pm 0.1$ as in Fig.~\ref{fig:ATLAS}. Again, we find that our calculations describe the data well for both centrality regions within the experimental error bars.

\section{Conclusions} \label{sec:conclusions}

In summary, we developed a new framework to describe the inclusive production of small-$R$ jets in heavy-ion collisions. For small-$R$ jets, the all order resummation of single logarithms of the jet radius parameter $\alpha_s^n\ln^n R$ has to be taken into account. We consistently calculated both the proton-proton and the heavy-ion jet cross section at NLO+NLL$_R$ accuracy. This new results was enabled by the use of the recently developed semi-inclusive jet functions. For the heavy-ion jet cross section, we introduced in-medium semi-inclusive jet functions analogously to the vacuum ones. We calculated the QCD medium contribution using the collinear in-medium splitting kernels derived within \SCETG to first order in opacity. We found good agreement with recent experimental measurements from the LHC for the nuclear modification factor $R_{AA}$ of inclusive jets. The calculation outlined in this work  sets the stage to address jet substructure observables in heavy-ion collisions in the future. (Semi-) inclusive jet substructure measurements performed on an inclusively measured jet in heavy-ion collisions can now be calculated consistently with the proton-proton baseline. See for example~\cite{Kang:2016ehg,Neill:2016vbi}.

\medskip

{\bf Acknowledgments}

We would like to thank Yang-Ting Chien, Raghav Elayavalli, Wouter Waalewijn and Hongxi Xing for helpful discussions. This research is supported by the US Department of Energy, Office of Science under Contract No. DE-AC52-06NA25396 and by the DOE Early Career Program under Grant No. 2012LANL7033.

\bibliographystyle{h-physrev5}
\bibliography{bibliography}

\begin{thebibliography}{10}

\bibitem{Wang:2016opj}
X.-N. Wang, editor,
\newblock {\em {Quark-Gluon Plasma 5}} (World Scientific, New Jersey, 2016).

\bibitem{Khachatryan:2016jfl}
CMS, V.~Khachatryan {\em et~al.},
\newblock Submitted to: Phys. Rev. C  (2016), arXiv:1609.05383.

\bibitem{Aad:2012vca}
ATLAS, G.~Aad {\em et~al.},
\newblock Phys. Lett. {\bf B719}, 220 (2013), arXiv:1208.1967.

\bibitem{Abelev:2013fn}
ALICE, B.~Abelev {\em et~al.},
\newblock Phys. Lett. {\bf B722}, 262 (2013), arXiv:1301.3475.

\bibitem{Sterman:1977wj}
G.~F. Sterman and S.~Weinberg,
\newblock Phys. Rev. Lett. {\bf 39}, 1436 (1977).

\bibitem{Cacciari:2008gp}
M.~Cacciari, G.~P. Salam, and G.~Soyez,
\newblock JHEP {\bf 04}, 063 (2008), arXiv:0802.1189.

\bibitem{Soyez:2008su}
G.~Soyez,
\newblock Nucl. Phys. Proc. Suppl. {\bf 191}, 131 (2009), arXiv:0812.2362.

\bibitem{Mukherjee:2012uz}
A.~Mukherjee and W.~Vogelsang,
\newblock Phys. Rev. {\bf D86}, 094009 (2012), arXiv:1209.1785.

\bibitem{Kang:2016mcy}
Z.-B. Kang, F.~Ringer, and I.~Vitev,
\newblock JHEP {\bf 10}, 125 (2016), arXiv:1606.06732.

\bibitem{Bauer:2000yr}
C.~W. Bauer, S.~Fleming, D.~Pirjol, and I.~W. Stewart,
\newblock Phys. Rev. {\bf D63}, 114020 (2001), arXiv:hep-ph/0011336.

\bibitem{Bauer:2001ct}
C.~W. Bauer and I.~W. Stewart,
\newblock Phys. Lett. {\bf B516}, 134 (2001), arXiv:hep-ph/0107001.

\bibitem{Bauer:2001yt}
C.~W. Bauer, D.~Pirjol, and I.~W. Stewart,
\newblock Phys. Rev. {\bf D65}, 054022 (2002), arXiv:hep-ph/0109045.

\bibitem{Beneke:2002ph}
M.~Beneke, A.~P. Chapovsky, M.~Diehl, and T.~Feldmann,
\newblock Nucl. Phys. {\bf B643}, 431 (2002), arXiv:hep-ph/0206152.

\bibitem{Dasgupta:2014yra}
M.~Dasgupta, F.~Dreyer, G.~P. Salam, and G.~Soyez,
\newblock JHEP {\bf 04}, 039 (2015), arXiv:1411.5182.

\bibitem{Dasgupta:2016bnd}
M.~Dasgupta, F.~A. Dreyer, G.~P. Salam, and G.~Soyez,
\newblock JHEP {\bf 06}, 057 (2016), arXiv:1602.01110.

\bibitem{Kaufmann:2015hma}
T.~Kaufmann, A.~Mukherjee, and W.~Vogelsang,
\newblock Phys. Rev. {\bf D92}, 054015 (2015), arXiv:1506.01415.

\bibitem{Dai:2016hzf}
L.~Dai, C.~Kim, and A.~K. Leibovich,
\newblock Phys. Rev. {\bf D94}, 114023 (2016), arXiv:1606.07411.

\bibitem{Vitev:2008rz}
I.~Vitev, S.~Wicks, and B.-W. Zhang,
\newblock JHEP {\bf 11}, 093 (2008), arXiv:0810.2807.

\bibitem{He:2011pd}
Y.~He, I.~Vitev, and B.-W. Zhang,
\newblock Phys. Lett. {\bf B713}, 224 (2012), arXiv:1105.2566.

\bibitem{Muller:2012zq}
B.~Muller, J.~Schukraft, and B.~Wyslouch,
\newblock Ann. Rev. Nucl. Part. Sci. {\bf 62}, 361 (2012), arXiv:1202.3233.

\bibitem{Zapp:2013zya}
K.~C. Zapp,
\newblock Phys. Lett. {\bf B735}, 157 (2014), arXiv:1312.5536.

\bibitem{Armesto:2015ioy}
N.~Armesto and E.~Scomparin,
\newblock Eur. Phys. J. Plus {\bf 131}, 52 (2016), arXiv:1511.02151.

\bibitem{Chien:2015hda}
Y.-T. Chien and I.~Vitev,
\newblock JHEP {\bf 05}, 023 (2016), arXiv:1509.07257.

\bibitem{Chang:2016gjp}
N.-B. Chang and G.-Y. Qin,
\newblock Phys. Rev. {\bf C94}, 024902 (2016), arXiv:1603.01920.

\bibitem{Wang:2016fds}
X.-N. Wang, S.-Y. Wei, and H.-Z. Zhang,
\newblock (2016), arXiv:1611.07211.

\bibitem{Idilbi:2008vm}
A.~Idilbi and A.~Majumder,
\newblock Phys. Rev. {\bf D80}, 054022 (2009), arXiv:0808.1087.

\bibitem{Ovanesyan:2011xy}
G.~Ovanesyan and I.~Vitev,
\newblock JHEP {\bf 06}, 080 (2011), arXiv:1103.1074.

\bibitem{Ovanesyan:2011kn}
G.~Ovanesyan and I.~Vitev,
\newblock Phys. Lett. {\bf B706}, 371 (2012), arXiv:1109.5619.

\bibitem{Fickinger:2013xwa}
M.~Fickinger, G.~Ovanesyan, and I.~Vitev,
\newblock JHEP {\bf 07}, 059 (2013), arXiv:1304.3497.

\bibitem{Ovanesyan:2015dop}
G.~Ovanesyan, F.~Ringer, and I.~Vitev,
\newblock Phys. Lett. {\bf B760}, 706 (2016), arXiv:1512.00006.

\bibitem{Kang:2016ofv}
Z.-B. Kang, F.~Ringer, and I.~Vitev,
\newblock (2016), arXiv:1610.02043.

\bibitem{Gyulassy:2000fs}
M.~Gyulassy, P.~Levai, and I.~Vitev,
\newblock Phys. Rev. Lett. {\bf 85}, 5535 (2000), arXiv:nucl-th/0005032.

\bibitem{Kang:2014xsa}
Z.-B. Kang, R.~Lashof-Regas, G.~Ovanesyan, P.~Saad, and I.~Vitev,
\newblock Phys. Rev. Lett. {\bf 114}, 092002 (2015), arXiv:1405.2612.

\bibitem{Chien:2015vja}
Y.-T. Chien, A.~Emerman, Z.-B. Kang, G.~Ovanesyan, and I.~Vitev,
\newblock (2015), arXiv:1509.02936.

\bibitem{Aversa:1988vb}
F.~Aversa, P.~Chiappetta, M.~Greco, and J.~P. Guillet,
\newblock Nucl. Phys. {\bf B327}, 105 (1989).

\bibitem{Jager:2002xm}
B.~Jager, A.~Schafer, M.~Stratmann, and W.~Vogelsang,
\newblock Phys. Rev. {\bf D67}, 054005 (2003), arXiv:hep-ph/0211007.

\bibitem{Aversa:1989xw}
F.~Aversa, M.~Greco, P.~Chiappetta, and J.~P. Guillet,
\newblock Z. Phys. {\bf C46}, 253 (1990).

\bibitem{Jager:2004jh}
B.~Jager, M.~Stratmann, and W.~Vogelsang,
\newblock Phys. Rev. {\bf D70}, 034010 (2004), arXiv:hep-ph/0404057.

\bibitem{Ritzmann:2014mka}
M.~Ritzmann and W.~J. Waalewijn,
\newblock Phys. Rev. {\bf D90}, 054029 (2014), arXiv:1407.3272.

\bibitem{Vogt:2004ns}
A.~Vogt,
\newblock Comput. Phys. Commun. {\bf 170}, 65 (2005), arXiv:hep-ph/0408244.

\bibitem{Anderle:2015lqa}
D.~P. Anderle, F.~Ringer, and M.~Stratmann,
\newblock Phys. Rev. {\bf D92}, 114017 (2015), arXiv:1510.05845.

\bibitem{Ellis:2010rwa}
S.~D. Ellis, C.~K. Vermilion, J.~R. Walsh, A.~Hornig, and C.~Lee,
\newblock JHEP {\bf 11}, 101 (2010), arXiv:1001.0014.

\bibitem{Collins:1988wj}
J.~C. Collins and J.-w. Qiu,
\newblock Phys. Rev. {\bf D39}, 1398 (1989).

\bibitem{Dulat:2015mca}
S.~Dulat {\em et~al.},
\newblock (2015), arXiv:1506.07443.

\bibitem{Vitev:2007ve}
I.~Vitev,
\newblock Phys. Rev. {\bf C75}, 064906 (2007), arXiv:hep-ph/0703002.

\bibitem{Kang:2016ehg}
Z.-B. Kang, F.~Ringer, and I.~Vitev,
\newblock JHEP {\bf 11}, 155 (2016), arXiv:1606.07063.

\bibitem{Neill:2016vbi}
D.~Neill, I.~Scimemi, and W.~J. Waalewijn,
\newblock (2016), arXiv:1612.04817.

\end{thebibliography}

\end{document}